# Using the AIM for solving the nonrelativistic wave equation for new class of infinite one dimensional well with none flat bottom

Ibsal A. Assi[a,1] Abdullah J. Sous[b], and Akpan N. Ikot[c]

[a] *Department of Physics and Physical Oceanography, Memorial University of Newfoundland, St. John's A1B3X7, NL, Canada*
[b] *Faculty of Technology and Applied Sciences, Al-Quds Open University, Tulkarm, Palestine*
[c] *Theoretical Physics Group, Department of Physics, University of Port Harcourt-Nigeria*

**Abstract**: The main goal of this work is to solve the nonrelativistic wave equation for a new potential configuration that describes the quantum states of a particle that lies within a one-dimensional infinite well of width $L$ using the Asymptotic Iteration Method (AIM). This potential was introduced recently by Alhaidari to be added to the class of exactly solvable potentials in the Tridiagonal Representation Approach (TRA). We have obtained the energy eigenvalues for different choices of the potential parameters. A good match between our results and the ones obtained by the TRA are shown in **Table 1**. Moreover, new results have been presented in **Table 2** and **Table 3** for the energy spectrum.

## 1. Introduction:

One of the main goals of quantum mechanics is to obtain solutions of the wave equation for as large class of potentials as possible. These solutions are important for better understanding of the corresponding physical systems. This topic took the attention of many scientists since the early days of quantum mechanics where authors used different analytical and numerical techniques to look for solutions of the relativistic and nonrelativistic wave equations. Among the methods that have been used we mention: Methods of super-symmetry [1], Nikiforov-Uvarov method [2], the Tridiagonal Representation Approach (TRA) [3], variational methods [4], the hypervirial perturbation method [5], and the shifted 1/N expansion techniques [6].

In this article, we will apply the Asymptotic Iteration Method (AIM) [7], to solve the nonrelativistic one-dimensional wave equation for the following new potential configuration introduced by Alhaidari [8]

$$V(x) = \begin{cases} \dfrac{A}{L^2 - x^2} + \dfrac{B}{x^2(L^2 - x^2)} + \dfrac{C}{(L^2 - x^2)^2} \; ; \; 0 < x < L \\ \infty \qquad\qquad\qquad\qquad\qquad\qquad ; x \leq 0 \,\&\, x \geq L \end{cases}, \quad (1.1)$$

---
[1] Corresponding Author: Email Address: iassi@mun.ca

where *A*, *B*, and *C* are constant potential parameters, and *L* is the width of the potential well. As an illustration, we plotted the potential in (1.1) for different choice of parameters in **Figure 1**. The potential in (1.1) has exact solutions in the TRA [3, 8], where the state wavefunction is written in terms of infinite bounded series in terms of Jacobi polynomials with expansion coefficients that represents new class of orthogonal polynomials that have not been studied before. The latter polynomials introduced by Alhaidari in [9]. However, the properties of those polynomials can be identified numerically at the moment and the hope is to determine their analytical properties in the near future in order to write the properties of the corresponding quantum system in closed form (e.g. energy spectrum).

The flow of our work goes as follows. In section 2, we present our review on the AIM. In section 3, however, we apply the AIM to our problem and give our results and discussions. Finally, we conclude our work in the fourth section.

## 2. Review on the Asymptotic Iteration Method:

In many physical problems, we end up having second order linear differential equation (SOLDE) that is needed to be solved to identify the properties of the associated physical system. The AIM deals with SOLDE of the following form [7]

$$f''(y) = \lambda_0(y) f'(y) + s_0(y) f(y), \qquad (2.1)$$

where $\lambda_0(y) \neq 0$, and $s_0(y)$ have sufficiently many continuous derivatives, i.e. $\lambda_0(y), s_0(y) \in C_\infty$. The symmetric structure of Eq. (2.1) implies the following differential equations of $f(y)$

$$f^{(n+1)}(y) = \lambda_{n-1}(y) f'(y) + s_{n-1}(y) f(y), \qquad (2.2)$$

and,

$$f^{(n+2)}(y) = \lambda_n(y) f'(y) + s_n(y) f(y), \qquad (2.3)$$

where,

$$\lambda_n = \lambda'_{n-1} + s_{n-1} + \lambda_0 \lambda_{n-1}, \quad s_n = s'_{n-1} + s_0 \lambda_{n-1}, \qquad (2.4)$$

The properties in (2.2) and (2.3) give

$$\frac{d}{dx} \ln f^{(n+1)}(y) = \frac{f^{(n+2)}(y)}{f^{(n+1)}(y)} = \frac{\lambda_n \left( y' + \frac{s_n}{\lambda_n} f(y) \right)}{\lambda_{n-1} \left( y' + \frac{s_{n-1}}{\lambda_{n-1}} f(y) \right)}, \qquad (2.5)$$

Now, if for sufficiently large *n* we have $s_n / \lambda_n = s_{n-1} / \lambda_{n-1} := \alpha(y)$, then Eq. (2.5) gives

$$f^{(n+1)}(y) = C_1 \exp\left( \int_0^y \frac{\lambda_n(t)}{\lambda_{n-1}(t)} dt \right) = C_1 \lambda_{n-1}(y) \exp\left( \int_0^y [\alpha + \lambda_0(t)] dt \right), \qquad (2.6)$$

Substituting (2.6) in (2.2) gives the following general solution of the SOLDE in (2.1)

$$f(y) = \exp\left(-\int_0^y \alpha(t)\,dt\right)\left[C_2 + C_1 \int_0^y \exp\left(\int_o^t (\lambda_0(u) + 2\alpha(u))\,du\right)dt\right], \qquad (2.7)$$

where physical restrictions require $C_1=0$. The last equation we need here is the termination condition (or the quantization condition) which reads

$$\Delta_n(y, E) = \lambda_n s_{n-1} - \lambda_{n-1} s_n = 0, \qquad (2.8)$$

The roots of Eq. (2.8) will give us the corresponding $n+1$ lowest energy eigenvalues of the system. In the next section, we will apply the AIM to the solve Schrodinger equation for the potential in (1.1).

## 3. Applying the AIM to the problem:

Our problem is described by the following one-dimensional Schrodinger equation (in units of $\hbar = m = 1$)

$$\left\{-\frac{1}{2}\frac{d^2}{dx^2} + \frac{A}{L^2 - x^2} + \frac{B}{x^2(L^2 - x^2)} + \frac{C}{(L^2 - x^2)^2}\right\}\psi(x) = E\psi(x), \qquad (3.1)$$

where $E$ is the energy eigenvalue, and $0 \leq x \leq L$. Before applying AIM to this problem, we have to transform the modified Schrödinger equation (3.1) to an amenable form for AIM. By using the change of variables, we can simplify Equation (3.1) by coordinate transformation $y = 2(x/L)^2 - 1$ which implies $y \in [-1, 1]$. Thus, Eq. (3.1) becomes

$$f''(y) = \lambda_0(y) f'(y) + s_0(y) f(y)), \qquad (3.2)$$

where

$$\lambda_0(y) = -\frac{1}{2(y+1)}, \qquad (3.3)$$

$$s_0(y) = -\frac{1}{4}\frac{EL^4 y^3 - 2EL^4 y^2 - EL^4 y + EL^4 + 2AL^2 y^2 - 2AL^2 + 4By - 4Cy - 4B - 4C}{L^2(y-1)^2(y+1)^2} \qquad (3.4)$$

We now find the first few functions of $\lambda_n(y)$ and $s_n(y)$ as,

$$\lambda_1(y) = \frac{3}{4(1-y)^2} + \frac{1}{4}\frac{EL^4 y^3 - EL^4 y + 2AL^2 y^2 + EL^4 - 2AL^2 + 4By - 4Cy - 4B - 4C}{L^2(y(y-1)-1)^4} \qquad (3.5)$$

$$s_1(y) = \frac{1}{4}\frac{3EL^4 y^2 - 4EL^4 y - EL^4 + 4AL^2 y + 4B - 4C}{L^2(y(y-1)-1)^4}$$

$$-\frac{(2y+1)(EL^4 y^3 - EL^4 y^2 + EL^4 - 2AL^2 + 4By - 4Cy - 4B - 4C)}{L^2(y(y-1)-1)^5}$$

$$+\frac{1}{4}\frac{EL^4 y^3 - EL^4 y^2 + 2AL^2 y^2 + EL^4 - 2AL^2 + 4By - 4Cy - 4B - 4C}{2L^2(y(y-1)-1)^4(1-y)} \qquad (3.6)$$

and so on. To calculate the eigen energies, we solve equation (2.8) for $E$. This will require us to choose certain values of $y$ such that we obtain fast convergence. Physically, the values of $E$ must be independent of our choice of $y$, usually different researchers' use the value of $y = y_0$ for which the potential has minimum or the point in which the ground state wave function takes its maximum value [7]. In principle, one can obtain a range of values of y such that we obtain stable result that are independent of our choice of $y_0$ within this range, which is commonly known as the *Plateau of Stability* (PoS). In this work, we took the midpoint of the $y$-space, that is $y_0 = 0$ to be the point of stability in the results as shown in **Table 1**, **Table 2** and **Table 3**. The bound state wave function is easily calculated using the following expression

$$\psi_n(y) = C_2 \left(1 - y\right)^a \left(1 + y\right)^b \exp\left(-\int_0^y \frac{s_n(t)}{\lambda_n(t)} dt\right), \tag{3.7}$$

where $y = 2(x/L)^2 - 1$.

## 4. Results and Discussion

As stated before the calculation of the energy eigenvalues $E_n$ by means of the quantization condition of the AIM depends on the choice of $y = y_0$ whose values lead to the speed of convergent of the eigenvalues as well as for the stability of the process. Now since the potential of equation (1.1) do not have exact analytical solution, we rather applied numerical methods to calculate the energy of equation (3.1) for the given sets of potential parameters. In this work we initialize $y_0 = 0$, corresponding to the midpoint of the $y$ interval. Thus, at the end of the iteration when a stable result is achieved, we obtain the energy spectrum by setting $y_0 = 0$. In **Table 1**, we compute energy eigenvalues $E_n$ using AIM and compare our results with TRA reported by Alhaidari for the following choice of potential parameters $A=4$, $B=4$, $C=8$, $L=2$. These results show good agreement with those reported by Alhaidari using TRA.

In **Table 2**, first column, we compute the energy level with the following potential values $A=0$, $B=4$, $C=8$, and $L=2$. In the second column, we considered $A=4$, $B=0$, $C=8$, and $L=2$. In a third column, we took $A=B=4$, $C=0$, and $L=2$. These are new results that have not been reported before. However, by simple comparison with the results of Table 1, we observed that the energy spectrum is strongly affected by the value of $B$. In contrast, by setting $A$ or $C$ equal to zero just made slight differences in the energy compared to the case when $B$ vanishes. Also in **Table 3**, we compute the lowest ten bound state energies for $A=-4$, $B=-4$, $C=-8$, $L=2$. These values of the parameters are just the negative of the values taken in Table 1. However, no comparison is made between the results of Tables 1 and 3 since the potential function in our case is not an odd function which indicates that our results make sense.

## 5 Conclusions

In this work we used AIM to calculate the energy spectrum for a new class of infinite one-dimensional well with none flat bottom. At the point of stability, we obtain the energy spectrum by setting $y = 0$ and these results are in good agreement with those reported in the literature using TRA.

Moreover, we would like to mention that this potential is less than 1 year old and never reported in literature to correspond to any physical problem. However, the hope is to find application(s) to this potential in physics.

## Figure Captions

**Figure 1**: Plot of the potential in (1) for different choices of parameters. In Figure 1(a), we took $A$, $B$, $C$= {{1,-2,3},{1,-2,18},{1,-2,27}}. In Figure 1(b), we took A, B, C= {{-3,-55,14},{-3,27,14},{-3,3.5,14}}.

## Table Captions

**Table 1**: A comparison of the energy eigenvalues $E_n$ obtained by the TRA and the AIM for the choice of potential parameters A=4, B=4, C=8, L=2.

**Table 2**: The lowest ten energy eigenvalues for different choices of parameters as indicated in each column.

**Table 3**: The lowest ten bound state energies for $A$=-4, $B$=-4, $C$=-8, $L$=2.

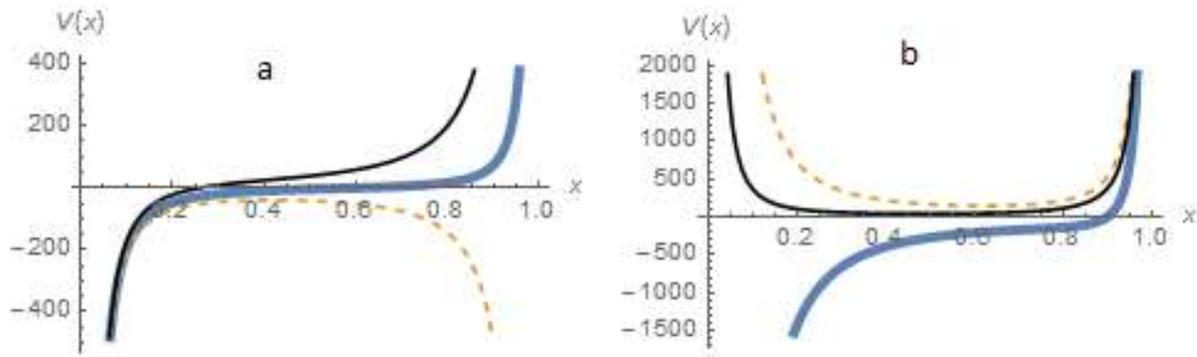

**Figure 1**

**Table 1**

| $n$ | $E_n$ (TRA) | $E_n$ (AIM) |
|---|---|---|
| 0 | 5.9727609687 | 5.972761814 |
| 1 | 11.9834081211 | 11.98341391 |
| 2 | 20.3733437263 | 20.37336112 |
| 3 | 31.1801373896 | 31.18018354 |
| 4 | 44.4219350633 | 44.42201870 |
| 5 | 60.1087189299 | 60.10888564 |
| 6 | 78.2465191684 | 78.24675783 |
| 7 | 98.8392406127 | 98.83966999 |
| 8 | 121.8895500217 | 121.8900467 |
| 9 | 147.3993465727 | 147.4004370 |

**Table 2**

| $n$ | $E_n$ [$A=0$, $B=4$, $C=8$, $L=2$] | $E_n$ $A=4$, $B=0$, $C=8$, $L=2$ | $E_n$ $A=B=4$, $C=0$, $L=2$ |
|---|---|---|---|
| 0 | 4.494939396 | 2.145125835 | 4.669130036 |
| 1 | 10.29011799 | 5.818782893 | 9.933633396 |
| 2 | 18.52601977 | 11.78962660 | 17.55700304 |
| 3 | 29.21393327 | 20.14801904 | 27.59113939 |
| 4 | 42.35926566 | 30.92900187 | 40.05865045 |
| 5 | 57.96507320 | 44.14916533 | 54.97104983 |
| 6 | 76.03310921 | 59.81732164 | 72.33491111 |
| 7 | 96.56475520 | 77.93893176 | 92.15431584 |
| 8 | 119.5604795 | 98.51723995 | 114.4319637 |
| 9 | 145.0222687 | 121.5547638 | 139.1697297 |

**Table 3**

| $n$ | $E_n$ |
|---|---|
| 0 | $-1.434082108 \times 10^5$ |
| 1 | -251.9142157 |
| 2 | -69.26639048 |
| 3 | -3.891328201 |
| 4 | -1.133144577 |
| 5 | 5.586514083 |
| 6 | 15.34323167 |
| 7 | 28.10647416 |
| 8 | 43.63009092 |
| 9 | 61.99556975 |